\documentclass[prd,a4paper,aps,epsfig,showpacs,twocolumn,floats]{revtex4}
\usepackage{epsfig}

\newcommand{\al}{\alpha}
\renewcommand{\b}{\beta}
\newcommand{\de}{\delta}

\newcommand{\ep}{\epsilon}

\newcommand{\la}{\lambda}
\newcommand{\Om}{\Omega}

\newcommand{\be}{\begin{equation}}
\newcommand{\ee}{\end{equation}}

\newcommand{\bea}{\begin{eqnarray}}
\newcommand{\eea}{\end{eqnarray}}
\newcommand{\bean}{\begin{eqnarray*}}
\newcommand{\eean}{\end{eqnarray*}}

\newcommand{\cf}{{{\cal F}}}
\newcommand{\ck}{{{\cal K}}}

\newcommand{\ar}{{A^r}}

\newcommand{\en}{{e^\nu}}

\newcommand{\ai}{{\alpha_i}}
\newcommand{\bi}{{\beta_i}}
\newcommand{\gi}{{\gamma_i}}
\newcommand{\mi}{{\mu_i}}
\newcommand{\nui}{{\nu_i}}
\newcommand{\ri}{{\rho_i}}
\newcommand{\tA}{{\tilde{A}}}
\newcommand{\tB}{{\tilde{B}}}
\newcommand{\tD}{{\tilde{D}}}
\newcommand{\tE}{{\tilde{E}}}
\newcommand{\rgm}{r_{gm}}

\begin{document}

\title{Generalized Einstein-Aether theories and the Solar System}

\author{Camille Bonvin$^{1,2}$, Ruth Durrer$^{2}$, Pedro G.~{\cal F}erreira$^{1}$, Glenn Starkman$^{1,3}$ and
Tom G.~Zlosnik$^{1}$}

\affiliation{
$^1$Astrophysics, University of Oxford, Denys Wilkinson Building, Keble 
Road, Oxford OX1 3RH, UK\\
$^{2}$ D\'epartement de Physique Th\'eorique, Universit\'e
de Gen\`eve, 24 quai Ernest Ansermet, CH--1211 Gen\`eve 4,
Switzerland.\\
$^{3}$ Department of Physics, Case Western Reserve University, Cleveland, OH 44106-7079}

\date{\today}

\begin{abstract}
It has been shown that generalized Einstein-Aether theories may lead to
significant modifications to the non-relativistic limit of the Einstein
equations. In this paper we study the effect of a general class of such
theories on the Solar System. We consider corrections to the gravitational potential in negative
{\it and} positive powers of distance from the source. Using measurements
of the perihelion shift of Mercury and time delay of radar signals to Cassini,
we place constraints on these corrections. We find that a subclass of
generalized Einstein-Aether theories are compatible with these constraints.
\end{abstract}


\maketitle

\section{Introduction}
\label{sec:intro}
The theory of general relativity proposed by Einstein explains a 
wealth of phenomena over a wide range of scales. At one extreme,
one obtains equations of motion for artificial satellites
in orbit around the Earth in particular and for Solar-System bodies
more generally that considerably surpass Newtonian gravity in the 
accuracy of their agreement with observations.
At the other extreme we find a comprehensive description
of the evolution of the Universe and 
of the growth of structure from a spectrum of primordial perturbations
which is backed up by a handful of precise astronomical observations. 
This apparent success over such a wide range of scales leads us to 
accept general relativity as the correct theory of gravitation 
valid on all scales.

It is well known that there are a number of imperfections in 
our understanding of the Universe, if we adopt Einstein's theory
as the theory of gravity and the Standard Model of particle physics
as the theory of matter.
For example, there is more gravity in galaxies and clusters of galaxies 
than can be accounted for by the matter (stars, gas, {\it etc.})
that we detect directly 
through absorption or emission of electromagnetic radiation \cite{DM}. 
This is explained by invoking the presence of dark matter, 
for which there are several reasonable candidates within
plausible extensions of the Standard Model of particle physics.
Furthermore, the Universe seems to be expanding in
a way that general relativity can only accomodate by
including an exotic form of energy that
possesses sufficient negative pressure to  be gravitationally repulsive
and that currently is the most  significant form of energy density on
cosmological scales. This is given the name dark energy 
if it varies in time or space, and the cosmological constant otherwise
\cite{DE}.
Finally, in order to understand the current homogeneity and isotropy
of the Universe and to provide an origin for the primordial
fluctuations  that grew into cosmic structures, we are
led to infer the existence of another such form of energy density
with negative pressure that dominated the evolution of the Universe
at early times and were the source of the quantum fluctuations which were the
progenitors of all structure.  This period of early dark-energy
domination is called inflation.
There are no particularly compelling candidates either for dark
energy or for inflation within extensions to the Standard Model 
not devised expressly for that purpose \cite{Inflation}.

As remarked, these imperfections can be explained by invoking the 
presence of dark matter and dark energy.
It is also conceivable that we do not yet
have the correct theory of gravity and that 
all or some of these effects are due to this fact.
This possibility has been the subject of intermittent attention
for a considerable time \cite{MOND,WeylGravity,MagBek}, 
but perhaps especially in the last few years \cite{DGP,fofR}.
In particular, there has been progress recently 
in devising covariant theories that can, apparently,
accommodate all these effects \cite{TeVeS,dodel,sanders,Tom,navvan}. 
At some level these are, consequently, promising alternative theories of gravity. 
As with general relativity, they must therefore satisfy the constraints
one infers from precise observations of the Solar System. 

In this paper we undertake the task of comparing one class of modified
gravity theories -- generalized Einstein-Aether theories -- 
to Solar-System observations.
In these theories, one assumes the existence of a 
vector field with a non-standard kinetic term, 
and with a time-like vacuum expectation value,
at least in the cosmological background.
Such theories lead to differences from general relativistic phenomenology
which may be substantial, for instance producing modifications to
Poisson's equation 
and the Friedmann equation of precisely the form posited as
alternatives to dark matter 
\cite{MOND} and dark energy \cite{DvaliTurner} respectively. They
should therefore be constrained by observations such as 
the perihelion precession of Mercury 
and the time delay of radio pulses around the Solar System.

The structure of the paper is as follows. 
We first review the Einstein-Aether theory in Section \ref{sec:field}, 
presenting the complete action and the field equations. 
We then focus on the spherically symmetric static metric in
Section \ref{sec:sphere} and consider a systematic expansion around
the Newtonian solution including terms that decrease more quickly with
distance as well as terms that grow with distance. By considering such
a wide range of solutions we find constraints on the parameters that define
the action presented in Section \ref{sec:field}. In Section
\ref{sec:conclusion} 
we discuss our results. We have organized the text in such a way that the main
thrust of the calculations are presented in the main body of the paper
while the details  are given in a set of appendices at the end.

\section{Field equations}
\label{sec:field}

A general action for a vector field $A^\al$ coupled to gravity has the following form: 
\begin{eqnarray}
S=\int d^4x \sqrt{-g}\left[\frac{R}{16\pi G_N}+{\cal L}(g,A)\right]
+S_{M}
\end{eqnarray}
where $g$ is the metric, $R$ the Ricci scalar and $S_M$ the matter action. 
The Lagrangian of the vector field ${\cal L}(g,A)$ is constructed to be both covariant and local. 
We use units with $c=\hbar=1$ and the metric signature is $(-,+,+,+)$. 
Furthermore we demand that the vector field is time-like with a fixed length $A^\al A_\al=-1$.

In this paper we will consider a Lagrangian that depends only on covariant derivatives of $A$ and the time-like constraint. 
It can be written in the form:
\begin{eqnarray}
\label{eq:Lagrangian}
{\cal L}(A,g)&=&\frac{M^2}{16\pi G_N}\left[
          {\cal F}({\cal K}) +\la(A^\alpha A_\alpha+1)\right] \\
{\cal K}&=&M^{-2}{\cal 
K}^{\alpha\beta}_{\phantom{\alpha\beta}\gamma\sigma}
\nabla_\alpha A^{\gamma}\nabla_\beta A^{\sigma} \nonumber \\
{\cal 
K}^{\alpha\beta}_{\phantom{\alpha\beta}\gamma\sigma}&=&c_1g^{\alpha\beta}g_{\gamma\sigma}
+c_2\delta^\alpha_\gamma\delta^\beta_\sigma+
c_3\delta^\alpha_\sigma\delta^\beta_\gamma \nonumber
\end{eqnarray}
where $c_i$ are dimensionless constants and $M$
has the dimensions of mass. $\lambda$ is a non-dynamical 
Lagrange-multiplier.

Note that in the particular case $c_1=-c_3$ and $c_2=0$ we recover the canonical form $\ck\propto F_{\al\beta}F^{\al\beta}$
where $\mathbf{F}$ is the field-strength of the four-vector $\mathbf{A}$.

The gravitational field equations for this theory are
\begin{equation}
\label{eq:einstein}
G_{\alpha\beta}=\tilde{T}_{\alpha\beta}+8\pi GT^{matter}_{\alpha\beta}
+M^2\lambda A_{\alpha}A_{\beta}
\label{fieldI}
\end{equation}
where the stress-energy tensor for the vector field is given by
\begin{eqnarray}
\tilde{T}_{\alpha\beta} &\equiv& \frac{1}{2}\nabla_{\sigma}
({\cal F}'(J_{(\alpha}^{\phantom{\alpha}\sigma}A_{\beta)}-
J^{\sigma}_{\phantom{\sigma}(\alpha}A_{\beta)}-J_{(\alpha\beta)}A^{\sigma}))
\nonumber \\ && -{\cal F}'Y_{\alpha\beta}
+\frac{1}{2}g_{\alpha\beta}M^{2}{\cal F}  ,
\end{eqnarray}
where
\begin{eqnarray}
{\cal F}' &\equiv& \frac{d{\cal F}}{d{\cal K}} \nonumber \\
J^{\alpha}_{\phantom{\alpha}\sigma} &\equiv&
2\cal{K}^{\alpha\beta}_{\phantom{\alpha\beta}\sigma\gamma}
    \nabla_{\beta}A^{\gamma}\\
Y_{\alpha\beta}&\equiv&c_{1}\left[(\nabla_{\al}A_{\sigma})(\nabla_{\beta}A^{\sigma})- (\nabla_{\sigma}A_{\al})(\nabla^{\sigma}A_{\beta})\right] . \nonumber
\end{eqnarray}
Brackets around indices denote symmetrization. 

The equations of motion for the vector field are
\begin{eqnarray}
\label{eq:motion}
\nabla_{\alpha}({\cal F}'J^{\alpha}_{\phantom{\alpha}\beta})
   &=& 2M^2\lambda A_{\beta}  .
\label{vectoreom}
\end{eqnarray}
Variation of the action with respect to $\lambda$ impose on the vector field the crucial constraint 
\be
\label{eq:constraint}
A^\al A_\al=-1 .
\ee

\section{Spherically symmetric static metric}
\label{sec:sphere}

In this paper we restrict ourselves to the particular case of a spherically symmetric static metric. 
In isotropic coordinates $(t,r,\theta,\phi)$, it takes the form:
\be
\label{eq:metric}
ds^2=-e^\nu dt^2+e^\xi(dr^2+r^2d\Om^2),
\ee
where $\nu(r)$ and $\xi(r)$  are functions of $r$ only.
The vector field has only two non-zero components: 
\be
A^\al=(A^t(r),A^r(r),0,0) .
\ee
With these restrictions, there are 5 unknown fields, $\nu, \xi, A^t,
A^r$ and the Lagrange multiplier $\la $, all of which are functions of
$r$ only. We need 5 equations to determine them. We choose the  $tt$
and $rr$ components of the Einstein equations  (\ref{eq:einstein}), 
the $t$ and $r$ components of the vector field  equation (\ref{eq:motion})
and  the constraint equation  (\ref{eq:constraint}). 
We can then combine these in order to eliminate the Lagrange
multiplier field $\la$,  leaving us with 3 dynamical
equations and the constraint.

The equations of motion depend on the function ${\cal F}(\ck)$ and 
its first and second derivatives with respect to $\ck$. In~\cite{Tom},
it was shown that  in order to get 'MONDian' corrections 
on galactic scales one has to choose the mass 
parameter $M$ to be of the order $M \sim H_0\sim 10^{-42}$ GeV. 
Therefore in the Solar System, where the gravitational field is strong with respect to $M$,  the function $\ck$ is much larger
than one, and $\cf'(\ck)$ can be expanded in inverse powers of $\ck^{1/2}$ 
\be
\label{eq:expand}
\cf'(\ck)=\sum_{i=1}^{\infty}\frac{\al_i}{\ck^{i/2}}~,
\ee
where this specific form is suggested by comparison to particular modifications of Poisson's equation
compatible with galaxy phenomenology \cite{Tom}. 
We therefore write the dynamical equations for a generic term $\cf'(\ck)=\frac{\al_{2n}}{\ck^n}$. 
Note that the leading-order term corresponds to $n=1/2$.  We find
\bea
\label{eq:t_tt}
\lefteqn{-\left(\xi''+2\frac{\xi'}{r}+\frac{\xi'^2}{4}\right)\frac{\ck^{n+1}}{\al_{2n}} =-\frac{M^2 e^\xi}{2}\cdot N(\ck)\ck}\\
&&+\Big(f_1\ck+nf_2\ck' \Big)\left(A^r \right)^2
+\Big(f_3\ck-nc_2e^\xi\ck'\Big)A^rA^{r}{}'\nonumber \\
&&+c_2e^\xi\ck\Big( (A^{r}{}')^2+A^rA^{r}{}''\Big)+\Big(f_4\ck+nf_5\ck' \Big)\left(A^t\right)^2\nonumber\\
&&+\Big(f_6\ck-nc_3e^\nu\ck' \Big)A^tA^{t}{}'+c_3e^\nu\ck\Big( (A^{t}{}')^2+A^tA^{t}{}''\Big) , \nonumber
\eea
\bea
\label{eq:r_rr}
\lefteqn{\left(\frac{\xi'+\nu'}{r}+\frac{\xi'^2}{4}+\frac{\xi'\nu'}{2}\right)\frac{\ck^{n}}{\al_{2n}} =\frac{M^2e^\xi}{2}\cdot
N(\ck)}\\
&&+g_1(A^r)^2 +g_2A^rA^{r}{}'-(c_1+c_2+c_3)e^\xi(A^{r}{}')^2\nonumber\\
&&+\frac{c_1-c_3}{2}(\nu{}')^{2}e^\nu(A^t)^2 \nonumber
+(c_1-c_3)\nu'e^\nu A^tA^{t}{}'\nonumber\\
&&+c_1e^\nu(A^{t}{}')^2 ,  \nonumber
\eea
and
\bea
\label{eq:t_r}
0&=&\Big(h_1\ck+nh_2\ck'\Big)A^rA^t+\Big(h_3\ck+2nc_1\ck'\Big)A^{t}{}'A^r\nonumber\\
&+&\Big(h_4\ck-2n(c_1+c_2+c_3)\ck'\Big)A^{r}{}'A^t\\
&+&2(c_1+c_2+c_3)\ck A^{r}{}''A^t-2c_1\ck A^{t}{}''A^r~,\nonumber
\eea
where 
\be
N(\ck)=\left\{ \begin{array}{ll}
\frac{\ck}{1-n} & \mbox{if $n\neq 1$}\\
\ln(\ck)\ck & \mbox{if $n=1$} 
\end{array}
\right.
\ee
and $f_i, g_i$ and $h_i$ are functions of $c_1, c_2, c_3$, the metric
fields and their first and second derivatives 
with respect to $r$. The specific expressions for these functions and
for $\ck$ in the metric (\ref{eq:metric}) are given in
appendix \ref{app:f_g_h}. 

Note that equation (\ref{eq:r_rr}) contains only first derivatives of
the metric and vector fields. It is therefore a constraint. For the 
leading order term at large $\ck$, $\cf'(\ck)=\frac{\al_1}{\sqrt{\ck}}$, 
the right-hand side vanishes and therefore this equation becomes  
\be
\label{eq:r_rr_0}
\left(\frac{\xi'+\nu'}{r}+\frac{\xi'^2}{4}+\frac{\xi'\nu'}{2}\right)=0~.
\ee

\subsection{Weak field approximation}

In the post-Newtonian parametrization, the metric fields are expanded
in power of $\frac{r_s}{r}$, where $r_s$ is the Schwarzschild radius
of the Sun. This expansion is a generalization of the asymptotic
behaviour of the Schwarzschild metric in isotropic coordinates far
from the source. Observations of the precession of Mercury and the
time delay of a signal emitted from the Earth and reflected by a
satellite or a planet allow us to constrain the first coefficients of
the expansion. Hence we try the post-Newtonian ansatz for the four fields  
\bea
\label{eq:expansion}
e^\nu&=&1+a_1\frac{r_s}{r}+a_2\left(\frac{r_s}{r}\right)^2+...\nonumber\\
e^\xi&=&1+b_1\frac{r_s}{r}+b_2\left(\frac{r_s}{r}\right)^2+...\\
A^r&=&d_1\frac{r_s}{r}+d_2\left(\frac{r_s}{r}\right)^2+...\nonumber\\
A^t&=&1+e_1\frac{r_s}{r}+e_2\left(\frac{r_s}{r}\right)^2+...~,\nonumber
\eea
where $a_i, b_i, d_i$ and $e_i$ are free coefficients which are determined by the equations of motions.

We shall take $A^{r}$ to have no constant component.
In fact,  there is a preferred cosmological frame in which
$A^{r}=0$, and the motion of the Sun with respect to that frame
will undoubtedly induce an $A^{r}$ \cite{jac_complet}.
Because the time-like vacuum expectation value of the aether
breaks Lorentz invariance, the effect of the solar motion with respect
to this frame cannot necessarily be accounted for by a boost,
however it suggests that the contribution to $A^{r}$ will be 
${\cal O}(\gamma_{sun}\beta_{sun})\sim 10^{-3}$.  We defer
further consideration of velocity-induced effects to  future work.

In order to recover Newton's theory we need $a_1<0$; the choice $a_1=-1$ defines Newton's constant.
Post-Newtonian corrections are associated with the parameters $a_2$ and $b_1$.
We insert the trial solutions (\ref{eq:expansion}) in the equations (\ref{eq:constraint}), (\ref{eq:t_tt}),
(\ref{eq:r_rr}) and (\ref{eq:t_r}), and we match the coefficients order by order in $\frac{r_s}{r}$ .

From equation (\ref{eq:t_tt}) at lowest order, we find: 
\be
\label{eq:c1}
n c_1=0.
\ee
The case $n=0$ corresponds to the standard aether theory with $\cf(\ck)=\ck$, and has already been studied
~\cite{jac_premier,jac_complet}. Here we are interested in $n\neq 0$ which therefore demands $c_1=0$.

It can be shown that $c_1=0$ leads to no modification to
Poisson's equation at all \cite{Tom}. 
Furthermore, it implies the existence of modes which propagate superluminally.
There is some debate as to whether this can be phenomenologically acceptable \cite{closed}.
We therefore try here to see whether the inclusion of additional terms 
in the expansion of the metric and vector field can
lead to a consistent weak field approximation for Lagrangians
with $c_1\neq 0$. 

\subsection{Additional terms}

In a modified theory of gravity, we expect the metric to have additional terms
to the ones given in (\ref{eq:expansion}). At the scale of a galaxy,
the next-to-leading term of the metric has to be a positive power or
logarithm of $r$ if the modified theory is to successfully mimic dark
matter on the scales of galaxies. More generally, unless there are
additional terms that at least decay more slowly than  $1/r$, 
the modified theory will have only minor phenomenological consequences
on large scales.
Since these additional terms have to be completely subdominant in the
Solar System, their contribution to the equations of motion have
traditionally been neglected. Nevertheless the form of the equations
(\ref{eq:t_tt}) and (\ref{eq:r_rr})  
suggests that they could play a role even in the Solar System. 
Indeed, in equation (\ref{eq:t_tt}) the left-hand side is proportional
to $\ck^{n+1}$, whereas the right-hand side is proportional to $\ck$. 
Since $\ck$ is expected to be much larger than one in the Solar System
\cite{Tom}, the parenthesis on the left-hand side must be
substantially suppressed.
Therefore even if the additional terms are small in the Solar System, 
especially because they appear on the left-hand side of equation
(\ref{eq:t_tt}) or (\ref{eq:r_rr}) they may measurably alter the
dynamic of the fields  in the Solar System. Indeed,
\cite{LueStarkmanDGP,LueStarkmanBirkhoff} show that, in modified
theories of gravity that seek to explain the observed accelerated
expansion of the Universe, 
corrections to the effective Newtonian potential of a massive body that grow with radius are generic.

Let us take the Minkowski metric as the background and add a spherically symmetric perturbation due to the Sun,
\bea
e^\nu&=&1+\phi(r),\\
e^\xi&=&1+\psi(r).\nonumber
\eea
We now expand the perturbations on a complete set, including both positive and negative powers of $r$: 
\bea
\label{eq:expansion_2}
\phi(r)&=&\sum_{i=1}^{\infty}a_i\left(\frac{r_s}{r}\right)^i +
\sum_{i=0}^{\infty}A_i\left(\frac{r}{\rgm}\right)^i \\
\psi(r)&=&\sum_{i=1}^{\infty}b_i\left(\frac{r_s}{r}\right)^i+
\sum_{i=0}^{\infty}B_i\left(\frac{r}{\rgm}\right)^i ~.\nonumber
\eea
Here $\rgm\equiv(r_s/M)^{1/2}$. 
One can show that this is the scale at which modifications of gravity occur   
in theories that attempt to modify the Einstein equations to
get MOND \cite{navvan}. 
For the Sun, $\rgm\simeq  10^{11}$ km. 
Since inside the Solar System $r\lesssim 10^9$ km, we have
$\frac{r}{\rgm}\lesssim  10^{-2}$.  

By a rescaling of the coordinates $t$ and $r$,
we can eliminate $A_0$ and $B_0$ such that the constant term is equal to 1. 
The coefficients in front of the additional terms, $A_i$ and $B_i$
can be constrained by observations in the Solar System.
Indeed,  observations of the perihelion shift of Mercury allow to
constrain the parameter $a_2$ to be $0.5$ with an accuracy of 
$\de a_2\simeq10^{-3}$, while time-delay observations allow us to constrain 
$b_1$ to be $1$ with an accuracy of $\de b_1\simeq10^{-5}$.

We derive limits on the coefficients $A_i$ and $B_i$ by requiring
 $|A_i|\left(\frac{r_M}{\rgm}\right)^i\lesssim \de
 a_2\left(\frac{r_s}{r_M}\right)^2$,  
where $r_M\simeq6\times 10^7$ km is the distance between Mercury and the Sun. 
The same argument applies for the other metric field. 
We find $|B_i|\left(\frac{r}{\rgm}\right)^i\lesssim \de b_1 \frac{r_s}{r}$.
The best constraint on $b_1$ comes from the Cassini satellite when it
 was at $\sim10^9$ km from the Sun ~\cite{will}.

These limits would be exact if the only non-zero term were the one under consideration. 
Otherwise, the true limits may be either weaker or stronger
(as the sum of a series can easily be smaller than some of its terms),
or even have the opposite sense  ($>$ instead of $<$ or vice versa).
For example, a correction  of the form $\exp(-C r/r_{gm})$ with $C>0$
requires a lower limit on $C$ rather than the upper limits
that would be derived from a term-by-term analysis.
Thus the limits we shall proceed to derive from a term-by-term approach
will be sufficient, but not necessary.  

Indeed, we show that  a set of consistency conditions between
the coefficients of the powers of $r/\rgm$  {\it can} be satisfied simultaneously. 
Thus an acceptable  solution exists which has an expansion that includes
both negative and positive powers of $r/\rgm$.
Future detailed study of the Schwarzschild metric in Einstein-Aether theories may yet 
identify an expansion basis that is better adapted to the physics
than polynomials in $r$.

One might also ask if these extra-terms in the metric are physically
acceptable if considered in isolation. After all, they diverge when $r$ goes to infinity. 
However, the expansion (\ref{eq:expand}) is itself appropriate only for 
$r \ll M^{-1} $, and in this paper we are looking to find solutions consistent with observational constraints in this range.
Meanwhile spatial infinity lies in the region $r \gg M^{-1}$, where another limiting form for ${\cal F}$ is appropriate.
We discuss the expected geometry in this region in Appendix \ref{secn:af}.

Through the equations of motion,
the perturbations of the metric created by the Sun lead to  perturbations of the vector field:
\bea
A^r(r)&=& \al (r) \\
A^t(r)&=&1+\beta(r) .\nonumber
\eea
We can again expand these perturbations on the complete set:
\bea
\label{eq:expansion_3}
\al(r)&=&\sum_{i=1}^{\infty}d_i\left(\frac{r_s}{r}\right)^i+\sum_{i=0}^{\infty}D_i\left(\frac{r}{\rgm}\right)^i
\\
\beta(r)&=&\sum_{i=1}^{\infty}e_i\left(\frac{r_s}{r}\right)^i+\sum_{i=0}^{\infty}E_i\left(\frac{r}{\rgm}\right)^i~.
\nonumber
\eea

In the following we restrict ourselves to the leading order term $n=1/2$ in the expansion (\ref{eq:expand}) for $\cf'$.
In appendix \ref{app:ordre_pert} we show that $\al(r)$ and $\beta(r)$ 
are of the same order-of-magnitude as the perturbations of the metric $\phi(r)$ and $\psi(r)$. 
Since the leading term of the metric in the Solar System is $\frac{r_s}{r}$, 
we will take this value at the edge of the Solar System as a limit for the peculiar terms. 
This means that 
\be
|D_i|\left(\frac{r}{\rgm}\right)^i\lesssim \frac{r_s}{r} \hspace{0.3cm} \textrm{and} \hspace{0.3cm}
|E_i|\left(\frac{r}{\rgm}\right)^i\lesssim \frac{r_s}{r}~.  
\ee
In table \ref{table}, we summarize the resulting constraints on the
coefficients $A_i, B_i, D_i$ and $E_i$ with $i\le 2$. 

\begin{table}
\begin{tabular}{|c | c | c|}
\hline
$e^{\nu}$&$e^{\xi}$&$A^t, A^r$  \\
\hline
$A_0=0$&$B_0=0$&$|D_0|,|E_0|\lesssim 3\cdot10^{-9}$ \\
\hline
$|A_1|\lesssim 10^{-14}$&$|B_1|\lesssim 10^{-11}$&$|D_1|, |E_1|\lesssim 10^{-6}$\\
\hline
$|A_2|\lesssim 10^{-10}$&$|B_2|\lesssim 5\cdot 10^{-9}$&$|D_2|,|E_2|\lesssim 5\cdot10^{-4}$\\
\hline
\end{tabular}
\caption{Constraints from observations on the coefficients of positive powers of $\frac{r}{\rgm}$.}\label{table}
\end{table}

Next we  wish to use our expansion of the metric
(\ref{eq:expansion_2}) and the vector field (\ref{eq:expansion_3}) to
solve equations (\ref{eq:t_tt}), (\ref{eq:r_rr}), (\ref{eq:t_r}) and
the constraint equation  (\ref{eq:constraint}) order-by-order in $r$.
We first rewrite the positive powers of $r$ in the form  
\be
A_n\left(\!\frac{r}{\rgm}\!\right)^n=A_n\left(\!\frac{r_s}{\rgm}\!\right)^n\left(\!\frac{r}{r_s}\!\right)^n\equiv 
\tA_n\left(\!\frac{r}{r_s}\!\right)^n~,
\ee
where $\tA_n\equiv A_n\left(\frac{r_s}{\rgm}\right)^n$. 
We do the same for the coefficients of the other fields: $B_n, E_n$ and $D_n$.
Since $\frac{r_s}{\rgm}\simeq 10^{-11}$, the constraints on the $A_n$ in 
table \ref{table} become much more stringent constraints on the $\tA_n$ 
(see table \ref{table2} in appendix \ref{app:positive}). 

We use a perturbative expansion in the coefficients. Moreover we
assume the hierarchy $1\gg |\tA_1|\gg |\tA_2|\gg ...$ for each of the four
fields in order to truncate the number of terms present at each order.  
The detailed calculation is presented in appendix~\ref{app:resolution}. 
Here we summarize the results.  

Equation (\ref{eq:t_r}) leads to two possibilities. 
The first is that $d_s=0$ and $\tD_s=0$ for all $s$, i.e. $A^r(r)=0$. 
This means that the only degree of freedom of the vector field is $A^t(r)$
which is then completely determined by the metric through the
unit time-like constraint. We do not study this specific
case in this paper, but rather concentrate on the other possibility
$A^r(r)\neq 0$. Nevertheless the former solution could 
be relevant in some
particular cases. For example in \cite{static}, it has been shown that
in the usual Einstein-Aether theory, corresponding to $n=0$, there
are regular perfect fluid stars with a static (i.e. $A^r(r)=0$) aether
exterior.

The choice $A^r(r)\neq 0$ implies that
the parameters $c_1, c_2$ and $c_3$ satisfy the constraint 
\be
(c_1+c_2+c_3)s^2-3(c_1+c_2+c_3)s-2(c_1-c_2+c_3)=0~,
\ee
for some positive integer $s$ which is such that $d_i=0$ for all $i<s$.

We have explicitly calculated the case $s=1$, which corresponds to
the constraint $c_3=-c_1$. From the positive orders in $\frac{r_s}{r}$
we find that $a_1$ is unconstrained. If we set $a_1=-1$ to recover
Newton's theory, we find   
\be
b_1=1+ O(10^{-9}) \hspace{0.5cm}\textrm{and}\hspace{0.5cm}
a_2=\frac{1}{2}+ O(10^{-9})~, 
\ee
which is in complete agreement with the observations which measure 
\be
b_1^{obs}=1\pm 10^{-3}  \hspace{0.5cm}\textrm{and}\hspace{0.5cm}
a_2^{obs}=\frac{1}{2}\pm 10^{-5}~. 
\ee
Moreover, we find that $c_1$ no longer has to be equal to zero. We have four 
solutions for $c_1$ and $c_2$ which are negative and therefore
causal~\cite{Tom}. They are given in appendix \ref{app:solution}. 
In each solution, $c_1$ and $c_2$ are of the order of
$\Big(\frac{\tB_2}{Mr_s}\Big)^2\frac{1}{\al^2_1}$. 
Since $\Big(\frac{\tB_2}{Mr_s}\Big)^2\sim 10^{-18}$, 
either $\al_1$ must be very small, or $c_1$ and $c_2$ are very small.

\begin{figure}[ht]
\centerline{\epsfig{figure=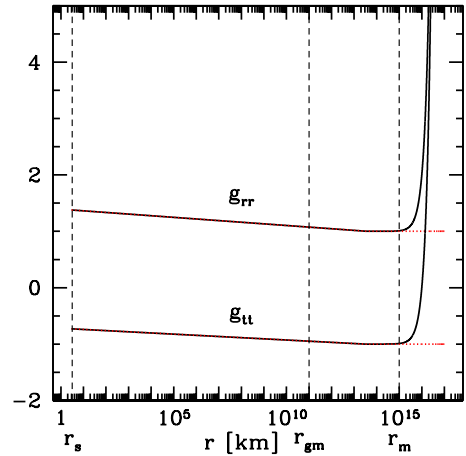,height=6.5cm}}
\caption{\label{fig1} The metric fields $g_{tt}$ and $g_{rr}$ as
  functions of $r$, in the aether theory (solid line) and 
in  general relativity (dotted line). We see that at the scale
$r_m\simeq10^{15}$ km the two models start to diverge. 
$r_{gm}\simeq 10^{11}$ km is the scale at which all growing
corrections are expected to start to dominate.}
\end{figure}

In figure \ref{fig1} we plot $g_{tt}(r)=-e^{\nu(r)}$ 
and $g_{rr}(r)=e^{\xi(r)}$ (solid line) 
for the values of $a_1$ to $a_4$, $b_1$ to $b_4$, $A_1$, $A_2$ and $B_1$ calculated in appendix \ref{app:sol_c1c3}. 
We find that $B_2$ is unconstrained by the equation of motion, except that $|B_2|< 5\cdot 10^{-9}$ (see table \ref{table}). 
We choose to saturate this maximum value for the plot. 
We also plot $g_{tt}$ and $g_{rr}$ from General Relativity, 
i.e. the Schwarzschild metric up to the order $\left(\frac{r_s}{r}\right)^4$ (dotted line). 
We see that inside the Solar System ($r\leq 10^9$ km) the two models are equivalent, 
but at a scale $r_m\simeq 10^{15}$ km they start to diverge. 
The Schwarzschild metric tends to $g_{tt}=-1$ and $g_{rr}=1$, whereas
the aether metric starts to grow. The scale $r_m$ is the one at
which the first two growing corrections ($r$ and $r^2$) become
dominant. The subsequent growing terms could start to grow
earlier. Nevertheless the limits set by hand on the coefficients $A_i$
and $B_i$ (see table \ref{table}) ensure that they cannot become
important inside the Solar System. Moreover $r_{gm}=10^{11}$ km is the
scale at which modifications of gravity should occur. Hence we expect
the growing terms to remain small for $r\lesssim r_{gm}$. Nevertheless,
 only the
full calculation of all the coefficients could confirm this limit.
Note that $g_{tt}(r)$ passes through zero at $r \simeq 10^{16}$ km and is therefore singular. However the series (\ref{eq:expansion_2}) cannot be taken seriously for $r>r_{gm}$, where they may not converge. 

Finally we have determined the structure of the equations at each order.
For negative powers in $r$ ($s>1$), 
\be
\label{eq:S}
S\left (\begin{array}{c} a_s\\b_s\\d_s\\e_s \end{array}\right)
=\left (\begin{array}{c} H_1^{(s)} \\H_2^{(s)}\\H_3^{(s)}\\H_4^{(s)} \end{array}\right)~, \hspace{0.7cm}\textrm{where}
\ee
\be
S\!\equiv\!\left(\!\begin{array}{cccc} -1&0&0&-2 \\0&s(s-1)&0&0\\-s&-s&0&0\\
\frac{d_1s(s-1)}{2}&0& -\frac{c_1(s-1)(s-2)}{4}&\frac{d_1s(s-1)}{2} \end{array}\!\right)
\ee
and $H^{(s)}_i$ are functions of the previous coefficients -- 
$a_1,...a_{s-1},b_1, ...,b_{s-1},d_1,..,d_{s-1},e_1,..,e_{s-1}$.
Therefore at order $s$ a unique solution $(a_s,b_s,d_s,e_s)$ exists if 
\be
\det S=-\frac{c_1}{2}s^2(s-1)^2(s-2)\neq 0~.
\ee
Since $c_1\neq 0$, this condition is satisfied for $s>2$. The order $s=2$, for which the determinant vanishes,
has been solved explicitly. The 
solutions are not unique and are given in appendix \ref{app:solution}. For each order $s>2$,
we can find the unique solution as a function of the
lower orders.
 
The same structure repeats for positive powers of $r$ ($s>2$):
\be
\label{eq:hatS}
\hat{S}\left (\begin{array}{c} \tA_s\\ \tB_s\\ \tD_s\\ \tE_s \end{array}\right)
=\left (\begin{array}{c} \hat{H}_1^{(s)}+\hat{Q}_1^{(s)} \\ \hat{H}_2^{(s)}+\hat{Q}_2^{(s)}\\\hat{H}_3^{(s)}+\hat{Q}_3^{(s)}
\\\hat{H}_4^{(s)}+\hat{Q}_4^{(s)} \end{array}\right)~, \hspace{0.7cm}\textrm{where}
\ee
\be
\hat{S}\!\equiv\!\!\left(\!\!\!\!\begin{array}{cccc} 1&\!\!\!\!\!\!\!\!\!0&\!\!\!\!\!\!\!\!\!0&2 \\0&\!\!\!\!\!\!\!\!-s(s+1)&\!\!\!\!\!\!\!\!\!0&0\\ s&\!\!\!\!\!\!\!\!\!s&\!\!\!\!\!\!\!\!\!0&0\\
\frac{c_1s(s+1)d_1}{2}&\!\!\!\!\!\!\!\!\!0&\!\!\!\!\!\! \!\!\!\begin{array}{c} [c_2(14-s)d_1^2-\\ \frac{c_1(s-1)(s-2)}{4}]\end{array} &\frac{c_1s(s+1)d_1}{2} \end{array}\!\!\!\!\right)
\ee
and $\hat{H}^{(s)}_i$ are functions of $a_1$, $b_1$, $d_1$ and $e_1$, 
and of the previous coefficients -- $\tA_1,...,\tA_{s-1}$, $\tB_1,...,\tB_{s-1}$,
$\tD_1,...,\tD_{s-1}$,$\tE_1, ...,\tE_{s-1}$.  In
appendix~\ref{app:positive} we argue that these functions are at
most of the same order of magnitude as the coefficient of order $s$.  
The $\hat{Q}_i^{(s)}$ are functions of the coefficients 
$\tA_{s+1}, \tA_{s+2},...$, $\tB_{s+1},...$,  $\tD_{s+1},...$, $\tE_{s+1},...$. 
Generically the $\hat{H}^{(s)}_i$ are much larger than the $\hat{Q}_i^{(s)}$ since the constraints
on the coefficients $\tA_s$ are much more stringent for larger $s$. We can, in general, therefore 
neglect the $\hat{Q}_i^{(s)}$. (See appendix \ref{app:positive} for a more detailed discussion).

A unique solution $(\tA_s, \tB_s, \tD_s, \tE_s)$ exists then at order $s$ if 
\be
\det \hat{S}=-2s^2(s+1)\Big(c_2(14-s)d_1^2-\frac{c_1}{4}(s-1)(s-2)\Big)\neq 0~.
\ee
The order $s=0, 1$ and 2 have been calculated in appendix \ref{app:solution}. For each of the solutions
(\ref{eq:solc1_1}) to (\ref{eq:solc1_4}) we find that $\det \hat{S}\neq 0~\forall ~s$.

The structure of equations (\ref{eq:S}) and (\ref{eq:hatS}) 
shows that 
there exist solutions to the equations of motion that can be expressed
as the expansions (\ref{eq:expansion_2}) for the metric and (\ref{eq:expansion_3}) for the vector field. 
Indeed, since at each order the determinants $\det S$ and $\det\hat{S}$ are non zero, we are  ensured that
a solution exists. We conclude that it is not sufficient to calculate the first two orders, 
even if we are only interested in the values of the Post-Newtonian parameters. 
In order to be sure that we have a {\it bona fide} solution of the equations of motion 
and that we will not encounter an inconsistency at any given order, we need to calculate
the determinant of the system at every order.

\section{Conclusion}

In this paper, 
we have studied the constraints we imposed from Solar System observations on generalized Einstein-Aether theory. 
We have considered an expansion of the metric around the Newtonian solution 
including the usual Post-Newtonian terms, which are negative powers of the distance $r$ from the Sun,
but also terms increasing with the distance (positive powers of $r$). 
The aim of this complete expansion is to take into account in the Solar System 
the effect of modification of gravity at large (Galactic and cosmological) scales. 
These effects are usually neglected in the Solar System 
and one considers only the decreasing terms in the metric expansion \cite{jac_premier,jac_complet,giannios}.  
Nevertheless, as long as the increasing terms are sufficiently small to be consistent with observations, 
nothing forces them to be completely absent in the Solar System.
Moreover in the case of generalized Einstein-Aether theory, we found that
the increasing terms play a crucial role, since without them the theory suffers from acausality. 
Indeed, if we neglect them, the equations of motion imply that one parameter of the theory $c_1=0$, 
leading to superluminal propagation of the aether field perturbations. If we consider the full expansion, the pathology disappears.

Moreover, we found that the Post-Newtonian parameters related to 
the precession of the perihelion of Mercury and the time delay of radio pulses are in agreement with observations. 
We found also constraints on the increasing terms coming from these observations, and constraints on the
parameters $c_1, c_2, c_3$ and $\al_1$ from the equations of motion. Indeed, whereas the $\alpha_{i}$ represent behaviour
in a particular regime of the theory (i.e. quasistatic configurations where the gravitational field is typically
much larger than the mass scale $M$), the $c_{i}$ are parameters of the Lagrangian which affect all solutions.
It may be shown \cite{Tom,Tom3} that simultaneous consideration of very weak gravitational fields,
the requirement for a realistic background cosmology, and suitable growth of large scale structure will
favour the $c_{i}$ to be O(1). We may immediately see then from equations (\ref{eq:solc1_1})-(\ref{eq:solc1_4})
that this corresponds to an upper limit on the number $\alpha_{1}$ of $10^{-9}$ to $10^{-8}$.
It is interesting to compare this bound to that obtained by considering the influence of such
a term on the Poisson equation then deducing the perihelion precession due to the extra force
provided by the modified potential in the context of Newtonian gravity. This was done in \cite{MOND}
where it was found that the resulting perihelion shift $\delta\phi_{\alpha}$ of Mercury's orbit per revolution, to lowest 
order in eccentricity, was given by:

\begin{equation}
\delta\phi_{\alpha}\sim 10\pi M\alpha_{1}\frac{r^{2}}{r_{s}}
\end{equation}

where $r$ is the semi-major axis length of Mercury's orbit. Taking $M=1.2\times 10^{-8}cm/s^{2}$, this yields a precession per
revolution which is approximately $\alpha_{1}/5$ of the prediction due to general relativity. Therefore, as observation
agrees with the prediction of general relativity to within $\sim 10^{-3}$, it was argued, $\alpha_{1}$ must
be less than around $5\times 10^{-3}$, therefore several orders of magnitude greater than our analysis will allow.
Clearly then, a benefit of recovering the modified Poisson equation from a set of generally covariant field
equations has been to allow a broadening of the scope of analysis of the theory's implications in the
solar system and in this case, in concert with additional constraints placed on the theory's parameters
in other regimes, this has allowed for a significantly more severe restrictions on the permissable size of such a modification.

Furthermore, we have developed a general method to test that the metric expansion is a solution of the equations of motion. The
constraints that we obtain must be interpreted, in part, as consistency
conditions on the theory.
Indeed, it is not sufficient to calculate the first coefficients in the expansion to confront the theory 
with Solar System observations, but one has to study carefully the structure of the equations at each order to be sure 
that no inconsistency will invalidate the results. Our results show that, even
though we have an infinite hierarchy of equations, these are solvable in
terms of the fundamental parameters of the theory. Hence, generalized
Einstein-Aether theories are viable theories of gravity within the Solar
System.

\label{sec:conclusion}

\acknowledgments
It is a pleasure to thank Chiara Caprini, Jean-Pierre Eckmann, Martin Kunz, Danail Obreschkow, Constantinos Skordis and Norbert Straumann
for useful and stimulating discussion. We also thank the referee for pointing out an erroneous statement in the
first version.
C. Bonvin is supported by the Swiss National Science Foundation and in part by a Royal Astronomical Society Grant.
R. Durrer is supported by the Swiss National Science Foundation.
T. Zlosnik is supported by a PPARC studentship.  G. Starkman was supported in part by the J.S. Guggenheim Memorial
Foundation, by the US DoE and by NASA.  Both CB and GDS thank Oxford Astrophysics for their hospitality.

\appendix

\section{On Asymptotic Flatness} \label{secn:af}

In the range $r \gg M^{-1}$ we expect the geometry to produce MOND behaviour which we will define as follows: We can decompose
the 4-velocity $U^{\mu}$ of a test particle into a part proportional to the velocity $C^{\mu}$ of a static observer
in the spacetime and a component $v^{\mu}$ satisfying $C^{\mu}v_{\mu}=0$. Normalization enforces
$U^{\mu}=(1+v^{2})^{\frac{1}{2}}C^{\mu}+v^{\mu}$ where
$v^{2}=g_{\mu\nu}v^{\mu}v^{\nu}$. The MOND regime is the regime for which
$v^{2}(r)=(r_{s}M/2)^{\frac{1}{2}}$. In the weak field limit
the constancy of $v^{2}$ at large radii can be interpreted as the levelling off of orbital velocity
for objects in circular orbits, for instance the so-called `flat rotation curves' of stars in approximately
circular orbits far from the centre of galaxies.

We would like then to obtain an insight into the expected geometry the model describes in this regime.
It can be shown that another relativistic theory which allows MOND, Bekenstein's TeVeS theory, can
be re-written in a form very similar to the model
discussed here \cite{tom1} i.e. it may be written as a theory with a single metric accompanied
by a derivative coupled Lorentz violating vector field with noncanonical kinetic terms. In TeVeS
for the static spherically symmetric case where the vector field points in the time direction
the metric for $r \gg M^{-1}$ takes the form in Schwarzschild coordinates $(t,\rho,\theta,\phi)$ \cite{skorzlo}:

\begin{equation}
\label{eqn:szmet}
ds^{2}=-\left(\frac{\rho}{\rho_{0}}\right)^{\frac{2n}{1-n}}dt^{2}+\frac{1}{(1-n)^{2}}d\rho^{2}+\rho^{2}d\Omega^{2}~, 
\end{equation}
where $\rho_{0}$ is a constant of integration,
$n\equiv \rho_{s}M/(4+\rho_{s}M)\ll 1$ and $\rho_s$ is the Schwarzschild radius of the Sun in Schwarzschild coordinates.
The form of $g_{00}$ is fixed,
via the geodesic equations, by the requirement that $v^{2}$ is independent of $\rho$. It can be shown that
this metric has nonvanishing components for the Riemann tensor
$R^{a}_{\phantom{a}bcd}$: $R^{t}_{\phantom{t}\phi t\phi} = (n-1)n\sin^{2}\theta,R^{\phi}_{\phantom{\phi}\theta\phi\theta} = -n(n-2),
R^{t}_{\phantom{t}\theta t \theta} = -n(n-1)$. It is this tensor, for instance, which is a measure of the change
$\Delta S_{a}$ in components of an arbitrary one-form $S_{a}$ parallel transported around a small closed
curve at some point $p$ in the spacetime \cite{wald}. Explicitly:

\begin{equation}
\Delta S_{b} = \frac{1}{2}R^{a}_{\phantom{a}bcd}S_{a}\oint x^{d}dx^{c}
\end{equation}

Clearly this is not generally zero if, as in this case, $R^{a}_{\phantom{a}bcd}$ does not vanish
and so we may conclude that the spacetime described by the above metric is not flat.  Due to the similarity between the
aether stress energy tensors in the two models, we expect the metric to take a very similar asymptotic
form for the model considered here when the aether points in the time direction. However, though
part of the geometry (and thus part of the necessary behaviour of the aether) is fixed by the requirement
that the MOND regime exists, a more thorough treatment is required to see how this is affected by allowing
for a radial component to the aether. It is interesting to note that for the case ${\cal F}={\cal K}$,
asymptotically flat solutions were only found for cases where the aether aligned with $\partial_{t}$ \cite{static}.
It is tempting then to conclude that for models with forms of ${\cal F}$ allowing a MOND regime
for $A \propto \partial_{t}$, the addition of radial terms, if allowing a MOND regime at all,
would not reduce the asymptotic curvature and that the absence of asymptotic flatness is generic. 

It may be checked that when (\ref{eqn:szmet}) is written in isotropic coordinates, the metric component
$g_{00}$ varies with the isotropic radial distance $r$ as $-r^{2n}$ whilst $g_{rr}$ varies
as $r^{-2n}$. The approach to the weak field limit of the metric can be found by requiring that $g_{00}$
is rather close to one, so we may expand $-(r/r_{0})^{2n}=-\exp(2n\ln(r/r_{0}))\simeq(1+2\ln(r/r_{0}))\equiv-(1+2\Phi)$.
The difference in the Newtonian potential $\Phi$ from the onset of MOND  at the typical scale of a galaxy
to the current Hubble radius is of the order $10^{-5}$ and so though the metric components do asymptotically
diverge, they are well within the weak field regime within the current cosmological horizon.
It is interesting then that the metric is approximately Minkowskian for $r\sim M^{-1}$ but diverges to greater curvature 
at far shorter \textit{and} far greater distances from the source.
However, the mildness of the deviation from flatness even on cosmological scales renders
it unclear whether the absence of asymptotic flatness of the kind in (\ref{eqn:szmet}) would
have observable consequences. It is also unclear whether models producing MOND and a vector
aligned with $\partial_{t}$ are more stable than solutions admitting a nonzero radial component
to the aether or whether they do tend to arise preferentially from astrophysical initial conditions.
Indeed it has been claimed that the static spherically symmetric solution with zero aether radial
component in TeVeS is unstable in the solar system with respect to certain perturbations,
the instability, partially manifest as an increasing radial component of the aether, growing
on unacceptably short time scales \cite{seif}. We postpone more detailed analysis to future work.

\section{Specific expression of the equations of motion}
\label{app:f_g_h}

The explicit expression for $\ck$ in the spherically symmetric and static metric is  
\bea
M^{2}\ck&=&(A^r)^2\Bigg[(c_1+c_3)\Big(\frac{2}{r^2}+\frac{2\xi'}{r}+\frac{\nu'^2}{4}+\frac{3\xi'^2}{4}\Big)\nonumber\\
&+&c_2\Big(\frac{4}{r^2}+\frac{2\nu'}{r}+\frac{6\xi'}{r}+\frac{3\nu'\xi'}{2} +\frac{\nu'^2}{4}+\frac{9\xi'^2}{4}\Big)\Bigg]\nonumber\\
&+&\Big[(c_1+c_3)\xi'+c_2\Big(\frac{4}{r}+\nu'+3\xi'\Big)\Big]A^rA^{r'} \\
&+&e^{\nu-\xi}(c_3-c_1)\Big[\frac{\nu'^2}{2}(A^t)^2 +\nu'A^tA^{t'}\Big]\nonumber\\
&+&(c_1+c_2+c_3)(A^{r'})^2-c_1e^{\nu-\xi}(A^{t'})^2~.\nonumber
\eea

The functions appearing in the equations of motions (\ref{eq:t_tt}), (\ref{eq:r_rr}) and (\ref{eq:t_r}) are given by:
\bea
f_1&=&\frac{e^\xi}{2}\Big[(c_1+c_3)\Big(\frac{\nu'^2}{2}+\frac{3\nu'\xi'}{2}+\frac{2\nu'}{r}+\nu''
\Big)~,\\
&& \hspace{-1.1cm} +c_2\Big(\frac{4}{r^2}+\frac{4(\nu'+3\xi')}{r}+
\frac{(\nu'+3\xi')^2}{2}+\nu''+3\xi'' \Big) \Big]~,\nonumber\\
f_2&=&-(c_1+c_3)\nu'-c_2\Big(\nu'+3\xi'+\frac{4}{r}\Big)~,\nonumber\\
f_3&=&e^\xi\Big[(c_1+c_3)\nu'+\frac{c_2}{2}\Big(3\nu'+9\xi'+\frac{12}{r} \Big)\Big]~,\nonumber\\
f_4&=&\frac{e^\nu}{2}(c_3-c_1)\Big(\frac{3\nu'^2}{2}+\frac{\nu'\xi'}{2}+\frac{2\nu'}{r}+\nu''\Big)~,\nonumber\\
f_5&=&-\frac{e^\nu}{2}(c_3-c_1)\nu'~,\nonumber\\
f_6&=&e^\nu\Big[\frac{c_3}{2}\Big(\frac{4}{r}+\xi'+5\nu'\Big)-c_1\nu'\Big]~,\nonumber\\
g_1&=&-\frac{e^\xi}{2}\Bigg[(c_1+c_3)\Big(\frac{\nu'^2}{2}+\frac{3\xi'^2}{2}+\frac{4\xi'}{r}+\frac{4}{r^2}\Big)~,\nonumber\\
&& \hspace{-1cm}+c_2\Big(\frac{8}{r^2}+\frac{\nu'^2}{2}+\frac{9\xi'^2}{2}+\frac{4(\nu'+3\xi')}{r}+3\nu'\xi'\Big) \Bigg]~,\nonumber\\
g_2&=&-e^\xi\Big[(c_1+c_3)\xi'+c_2\Big(\nu'+3\xi'+\frac{4}{r} \Big)\Big]~,\nonumber\\
h_1&=&-c_1\Big(\nu'^2+\frac{4}{r^2}+\frac{2(\nu'+\xi')}{r}+\nu''-\xi''\Big)~,\nonumber\\
&&+c_3\Big(\nu'\xi'-\frac{4}{r^2}+\frac{2(\nu'-\xi')}{r}+\nu''+\xi''\Big)~,\nonumber\\
&&+c_2\Big(\nu''+3\xi''-\frac{4}{r^2}\Big) ~,\nonumber\\
h_2&=&c_1(\nu'-\xi')-c_2\Big(\nu'+3\xi'+\frac{4}{r}\Big)-c_3(\nu'+\xi')~,\nonumber\\
h_3&=&-c_1\Big(3\nu'+\xi'+\frac{4}{r}\Big)~~\textrm{and}\nonumber\\
h_4&=&(c_1+c_2+c_3)\Big(\nu'+3\xi'+\frac{4}{r}\Big)\nonumber
\eea

\section{Order of magnitude of the vector field perturbations}

\label{app:ordre_pert}

In this appendix, we compare orders of magnitude of the different field perturbations. We know that in the Solar System the
metric perturbations are of the order $\phi(r)\simeq\psi(r)\simeq \frac{r_s}{r}$.  We want to calculate the order of magnitude of the
vector field perturbations $\alpha(r)$ and $\beta(r)$.

At first order in the perturbations:
\bea
\nu'&=&\frac{\big(e^\nu\big)'}{e^\nu}\simeq \phi'~,\hspace{0.7cm} \xi'\simeq \psi'~,\\
A^{t'}&=& \beta'~~\textrm{and} \hspace{0.7cm} A^{r'}=\alpha'~.\nonumber
\eea
The constraint (\ref{eq:constraint}) gives 
\be
-(1+\phi)(1+\beta)^2+(1+\psi)\al^2=-1~.
\ee
At the lowest order in each perturbation we have
\be
-\phi-2\beta+\al^2=0~.
\ee

We have four possibilities:
\begin{enumerate}
\item The three terms are or the same order of magnitude
  $\beta\sim\al^2\sim\phi$ which implies $\al^2=\phi+2\beta$
  \label{same}; 
\item $\al^2\ll\phi\sim\beta$ which implies $\phi\simeq -2\beta$;
\item $\beta\ll\phi\sim\al^2$ which implies $\al^2\simeq \phi $;\label{imaginaire}
\item $\phi\ll\beta\sim\al^2$ which implies $\al^2\simeq 2\beta$ \label{vector}.
\end{enumerate}
We see directly that case \ref{imaginaire} is not possible, since
$\phi= -r_s/r<0$ at first order in the Solar system. We use
equation (\ref{eq:t_r}) to exclude the possibilities \ref{same} and
\ref{vector}. Indeed, in these two cases
we have $\al\gg \beta,\phi$. Using these constraints we find 
\bea
\ck=\frac{1}{M^2}\Big[\frac{2\al(c_1+2c_2+c_3)}{r^2}&+&\frac{4\al'\al c_2}{r}\\
&+&(c_1+c_2+c_3)\al'^2 \Big]~,\nonumber
\eea
and equation (\ref{eq:t_r}) becomes
 \bea
0&=& 2(c_1+c_2+c_3)\Big[-2\frac{\al}{r^2}+2\frac{\al'}{r}+\al'' \Big]\ck\nonumber\\
&& -\Big[c_2\frac{\al}{r}+(c_1+c_2+c_3)\al'\Big]\ck'=0~.
 \eea
 
The only solution of this equation is $\al(r)=0$, which implies that
equation (\ref{eq:t_r}) doesn't allow the spatial component 
$A^r$ to be much larger than the other perturbations. So the only possibility is that the dominant contribution to the time component of the 
vector field is $\beta=-\frac{r_s}{2r}$, and that the dominant contribution to the space component satisfies $\al^2\ll \frac{r_s}{r}$. So it
is legitimate to assume that also  $\al\sim \frac{r_s}{r}$ at first order. 

\section{Resolution}

\label{app:resolution}

In this appendix we present the detailed derivation of the
solutions. The aim is to take the expansion (\ref{eq:expansion_2}) for
the metric and (\ref{eq:expansion_3}) for the vector field, to insert
it in the equations of motions and to solve order by order in power of 
$r$, using a perturbative expansion for the positive powers of
$r$. The orders $r^{-1}$ and $r^{-2}$ give the values of the
post-Newtonian parameters $b_1$ and $a_2$. The orders $r^{-3}, r^{-4}$
and $r^0$, $r^1$ and $r^2$ allow us to constrain the values of the
$c_i$ which are compatible with the data. Nevertheless it is not
sufficient to calculate only these orders. Indeed, each equation
generates an infinite numbers of positive and negative orders, and the
expansions (\ref{eq:expansion_2}) and (\ref{eq:expansion_3}) are
solutions of the equations of motion only if each of these equations
has a solution. Therefore, it is crucial to understand the structure of
the equations at each order and to verify that one introduces a sufficient
number of new coefficients at each order so that the equations do not
lead to constraints between coefficients which have 
already been determined, leading to a possible inconsistency.

In \ref{app:negative}, we consider only the terms with negative powers
of $r$, that means the terms which appear in the usual Post-Newtonian
parametrization . In~\ref{app:positive} we will consider the positive
powers. Finally, in~\ref{app:modification} and \ref{app:solution} we
apply our result to equations~(\ref{eq:t_tt}), (\ref{eq:r_rr}), 
(\ref{eq:t_r}) and (\ref{eq:constraint}).

\subsection{Negative powers of $r$}
\label{app:negative}

We have four fields $A^r, A^t, e^\xi$ and $e^\nu$, satisfying four equations. Three fields, $A^t, e^\xi$ and $e^\nu$  have 
an expansion of the form 
\be
\label{eq:exp_1}
e^\nu=1+\sum_{n=1}^{\infty}a_nx^n~,
\ee
where $x=\frac{r_s}{r}$.
The other field has no constant term in the expansion 
\be
\label{eq:exp_2}
A^r=\sum_{n=1}^{\infty}d_nx^n~.
\ee 

In the following we restrict ourselves to the two fields: $\ar$ and
$\en$ and two equations. 
The generalization for the four fields will then be straightforward.

We assume that the two equations governing the two fields have the form 
\be
\label{eq:simple_form}
\sum_{i=1}^{m}f_i r^{\sigma_i}(\en)^\ai(\en')^\bi(\en'')^\gi(\ar)^\mi(\ar')^\nui(\ar'')^\ri=0~,
\ee
where a$'$ stands for $\frac{d}{dr}$ and the exponents are all
positive integers or null, except $\sigma_i$ which is negative.
Actually $\ai$ will in principle be negative, since $\nu'=(\en)'(\en)^{-1}$ 
and $\nu''=(\en)''(\en)^{-1}-(\en')^2(\en)^{-2}$, but we can always 
multiply by $(\en)^j$ so that $\ai$ is a positive integer or null for all $i$. 

The form (\ref{eq:simple_form}) is not exactly the one which appears
in equations (\ref{eq:t_tt}) and (\ref{eq:r_rr}), 
because of the term $K^n$, where $n$ is an integer or a
half integer. Nevertheless, we will see later that we can rewrite the
problem in such a way that the equation is the form~(\ref{eq:simple_form}).

Since all the terms in the sum must have the same dimension, we have that $-\sigma_i+\bi+2\gi+\nui+2\ri=c$ is the same for all $i$.
Therefore, we can solve for $\sigma_i$, and then multiply every term by $r^c$. We insert then the expansion (\ref{eq:exp_1}) and
(\ref{eq:exp_2}), and their derivatives to obtain 
\bea
\label{eq:simple_form_dev}
\lefteqn{\sum_{i=1}^{m}f_i(-1)^{\bi+\nui}x^{-(\bi+2\gi+\nui+2\ri)}\left(1+\sum_{n=1}^{\infty}a_nx^n\right)^\ai}\nonumber\\
\cdot&&\left(\sum_{n=1}^{\infty}n a_n x^{n+1}\right)^\bi
\left(\sum_{n=1}^{\infty}n(n+1)a_n x^{n+2}\right)^\gi\nonumber\\
\cdot&&\left(\sum_{n=1}^{\infty}d_nx^n\right)^\mi \left(\sum_{n=1}^{\infty}n d_n x^{n+1}\right)^\nui\nonumber\\
\cdot&&\left(\sum_{n=1}^{\infty}n(n+1)d_n x^{n+2}\right)^\ri=0~.
\eea
Each term of the product can be expanded in power of $x$ using the binomial theorem
\be
\big(e^\nu\big)^{\al_i}=\left(1+\sum_{n=1}^{\infty}a_n x^n\right)^\ai=1+\ai a_1x +...
\ee
\bea
\big(A^r\big)^{\mi}&=&\left(\sum_{n=1}^{\infty}d_n x^n\right)^\mi\\
&=&d_1^\mi x^\mi + \mi d_1^{\mi-1}d_2 x^{\mi+1} + ... \nonumber\\
&& \hspace{-1cm} +\Big\{\mi d_1^{\mi-1}d_s + g_s(d_1,\cdots,
d_{s-1})\Big\} x^{\mi+s-1}+...\nonumber
\eea
where $g_s(d_1,\cdots, d_{s-1})$ is a function of the coefficients of order 1 to $s-1$. 
\bea
\big(e^{\nu'}\big)^{\beta_i}&=&\left(\sum_{n=1}^{\infty}n a_n x^{n+1}\right)^\bi\nonumber\\
&&=d_1^\bi x^{2\bi}  +\cdots
\eea
and the same for $\ar''$.

We can insert these developments in equation (\ref{eq:simple_form_dev}) and solve order by order in $x$.
The lowest order is 
\be
\sum_{i\in O_1}f_i(-1)^{\bi+\nui}2^{\gi+\ri}a_1^{\bi+\gi}d_1^{\mi+\nui+\ri}x^{\bi+\gi+\mi+\nui+\ri}=0~,
\ee 
where $O_1=\Big\{i\in\{1,...,m\} ~\textrm{such that}~
\bi+\gi+\mi+\nui+\ri=\delta ~\textrm{is the smallest exponent of the
  sum}\Big\}$. 

The next order in $x$ contains two different contributions. The first
contribution comes from the terms in $O_1$ where we take the next order in
the expansion for one term of the product, and the lowest order for
the others. This contribution contains two new coefficients $a_2$ 
and $d_2$ which appear linearly, and also the previous coefficients $a_1$ and $d_1$.
The second contribution comes from other terms in the sum, $i\in O_2$, where $O_2=\Big\{i\in\{1,...,m\} ~\textrm{such that}~
\bi+\gi+\mi+\nui+\ri=\delta+1\Big\}$. This contribution contains only $a_1$ and $d_1$. The next order $x^{\delta+1}$ is therefore 
\bea
\lefteqn{\hspace{-0.9cm}\sum_{i\in O_1}f_i(-1)^{\bi+\nui}2^{\gi+\ri}\Big\{(2\bi+3\gi)a_1^{\bi+\gi-1}d_1^{\mi+\nui+\ri}a_2}\nonumber\\
&&+(\mi+2\nui+3\ri)a_1^{\bi+\gi}d_1^{\mi+\nui+\ri-1}d_2\Big\} x^{\delta+1}\nonumber\\
&&+H^{(1)}_1(a_1,d_1)x^{\delta+1}=0~.
\eea
More generally, at the order $x^{\delta+s-1}$ ($s>1$) the two equations have the form 
\bea
\label{eq:order_s}
\lefteqn{F^{(s)}_1(a_1,d_1)\cdot a_s+G^{(s)}_1(a_1,d_1)\cdot d_s} \nonumber\\
&&=H^{(s)}_1(a_1,\cdots, a_{s-1}, d_1,\cdots, d_{s-1})\,, \nonumber\\
\lefteqn{F^{(s)}_2(a_1,d_1)\cdot a_s+G^{(s)}_2(a_1,d_1)\cdot d_s} \nonumber\\
&&=H^{(s)}_2(a_1,\cdots, a_{s-1}, d_1,\cdots, d_{s-1})\,,
\eea
where
\bea
\lefteqn{F^{(s)}_1(a_1,d_1)=\sum_{i\in O_1}f_i(-1)^{\bi+\nui}2^{\gi+\ri-1}\cdot}\nonumber\\
&\times&\Big(2s\bi+s(s+1)\gi\Big)a_1^{\bi+\gi-1}d_1^{\mi+\nui+\ri}\nonumber\\
\lefteqn{G^{(s)}_1(a_1,d_1)=\sum_{i\in
    O_1}f_i(-1)^{\bi+\nui}2^{\gi+\ri-1}} \nonumber\\
&\times&\Big(2\mi+2s\nui+s(s+1)\ri\Big)a_1^{\bi+\gi}d_1^{\mi+\nui+\ri-1}
\nonumber \,, \\
\eea
and $H^{(s)}_1(a_1,\cdots, a_{s-1}, d_1,\cdots, d_{s-1})$ is a function of the coefficients of order less than $s$.
$F^{(s)}_2(a_1,d_1)$ and $G^{(s)}_2(a_1,d_1)$ have the same form but with different $f_i$ and $O_1$.

A unique solution $(a_s, d_s)$ exists at order $s$ if 
\be
\label{eq:det}
\det \left(\begin{array}{cc}F^{(s)}_1(a_1,d_1)&G^{(s)}_1(a_1,d_1)\\ F^{(s)}_2(a_1,d_1)&G^{(s)}_2(a_1,d_1) ) \end{array}\right)\neq 0~.
\ee

Since the functions $F^{(s)}_j$ and $G^{(s)}_j$ contain only $a_1$,
$d_1$ and the parameters $f_i$ for $i\in O_1$, it is easy to calculate them at each order, without solving the entire system.
We only have to solve explicitly the first order. Then if the determinant is non zero for all $s>1$
we know that a unique solution exists which can be determined as a function of the previous coefficients. On the other hand if 
the determinant vanishes at some order $s$, an inconsistency may occur. Indeed, equations (\ref{eq:order_s}) can become a constraint
between the previous coefficients and lead to a contradiction. Therefore, if one wants to test the validity of a theory
in the PPN parametrization, it is not sufficient to calculate the first coefficients in the expansion and to compare
them with the observations. One has also to
calculate the determinant (\ref{eq:det})  at each order $s$, to insure that 
no additional constraints on the first coefficients will occur from higher orders.

The generalization of this method to four equations with four fields is straightforward. Each field with no constant component generates 
a function of the form $F^{(s)}_j$, whereas fields with a constant component generate a function of
the form $G^{(s)}_j$.

\subsection{Positive powers of $r$}
\label{app:positive}

In this section, we add positive powers of $r$ to the expansions (\ref{eq:exp_1}) and (\ref{eq:exp_2}):
\bea
e^\nu&=&1+ \sum_{n=1}^{\infty}A_n\left(\frac{r}{\rgm}\right)^n+ \sum_{n=1}^{\infty}a_nx^n\nonumber\\
&=&1+ \sum_{n=1}^{\infty}A_n\ep^n \frac{1}{x^n}+ \sum_{n=1}^{\infty}a_nx^n~,\nonumber\\
\eea
where $\ep=\frac{r_s}{\rgm}\simeq 10^{-11}$.
Equivalently for $A^r$ 
\be
A^r= \sum_{n=0}^{\infty}D_n\ep^n \frac{1}{x^n}+ \sum_{n=1}^{\infty}d_nx^n~.
\ee

These new terms add an infinite number of contributions to each
previous order $x^{\delta+s}$. Furthermore they generate new lower orders $x^{\delta-s}$. To simplify the problem we must take into account
the fact that the coefficients in front of the negative orders $A_n\ep^n$ (respectively $D_n\ep^n$) have to be small in order not to be observed in the Solar System.
Indeed the constraints  of table \ref{table} are equivalent to
$|A_n|\ep^n\lesssim \de a_2 \left(\frac{r_s}{r_M}\right)^n\ll 1$. 
Therefore, we use a perturbative expansion for $A_n\ep^n$ and $D_n\ep^n$. Furthermore since $\frac{r_s}{r_M}\simeq 5\cdot 10^{-8}\ll 1$
we have the following hierarchy: 
$1\gg |A_1|\ep \gg |A_2|\ep^2...$.  (And similarly for the $D_n\ep^n$) So we start studying the effect of $D_0$
(remember that $A_0$ has been put to zero by a rescaling of $t$ and $r$), then $\frac{A_1\ep}{x}$ and $\frac{D_1\ep}{x}$ and so on.
And we neglect all these new terms
with respect to 1. In other words, we take them into account only when they introduce a new order to the equation $x^{\delta-s}$.

Let us define new coefficients to simplify the formulae.
\be
\tA_n=\ep^nA_n\hspace{1.3cm}\tD_n=\ep^nD_n~.
\ee

In terms of these coefficients, the constraints coming from
observations are as follows:

\begin{table}[ht]
\begin{tabular}{|c | c | c|}
\hline
$e^{\nu}$&$e^{\xi}$&$A^t, A^r$  \\
\hline
$\tA_0=0$&$\tB_0=0$&$|\tD_0|,\, |\tE_0|\lesssim 3\cdot 10^{-9}$ \\
\hline
$|\tA_1|\lesssim 10^{-25}$&$|\tB_1|\lesssim 10^{-22}$&$|\tD_1|,\,
|\tE_1|
\lesssim 10^{-17}$\\
\hline
$|\tA_2|\lesssim 10^{-32}$&$|\tB_2|\lesssim
5\cdot10^{-31}$&$|\tD_2|,\, |\tE_2|\lesssim \cdot 10^{-26}$\\
\hline
\end{tabular}
\caption{Constraints from observations on the new coefficients .}\label{table2}
\end{table}

\subsubsection{Effect of $\tD_0$} 

\be
\left(\tD_0 +\sum_{n=1}^{\infty}d_n x^n\right)^\mi= \tD_0 d_1^{\mi-1} x^{\mi-1}+ d_1^\mi x^\mi + \cdots O(\tD_0^2)
\ee
The term $\tD_0$ introduces a new order $x^{\delta-1}$ 
\be
\sum_{i\in O_1}f_i(-1)^{\bi+\nui}2^{\gi+\ri}a_1^{\bi+\gi}d_1^{\mi+\nui+\ri-1}\tD_0 \cdot x^{\delta-1}
\ee

\subsubsection{Effect of $\tA_1 x^{-1}$ and $\tD_1 x^{-1}$}

At first order in $\tA_1$ and $\tD_1$ we have 
\bea
\left(\en \right)^\ai &=&
\left(1+\frac{\tA_1}{x}+\sum_{n=1}^{\infty}a_n x^n\right)^\ai
\nonumber \\
  &=&\ai\frac{\tA_1}{x}+1+\ai a_1x +\cdots
\eea
\bea
\left(\ar \right)^\mi&=&\left(\frac{\tD_1}{x}+\sum_{n=1}^{\infty}d_n x^n\right)^\mi=\mi \tD_1 d_1^{\mi-1}x^{\mi-2}\nonumber\\
&+&\mi(\mi-1) \tD_1 d_1^{\mi-2}d_2 x^{\mi-1} + d_1^\mi x^\mi +\cdots \nonumber\\
\eea
\bea
\lefteqn{\left(\en' \right)^\bi=\left(\tA_1-\sum_{n=1}^{\infty}n a_n x^{n+1}\right)^\bi=} \nonumber\\
&&(-1)^{\bi-1}\frac{\bi(\bi-1)}{2} \tA_1 a_1^{\bi-1}
x^{2\bi-2}\nonumber\\
&+&(-1)^{\bi-1}\bi(\bi-1)^2 \tA_1 a_1^{\bi-2}a_2 x^{2\bi-1}
\nonumber\\
&& + (-1)^\bi \bi a_1^\bi x^{2\bi} +\cdots \,, \nonumber\\
\eea
and equivalently for $\ar'$. $\en''$ and $\ar''$ remain the same as
previously. Therefore, we see that $\tA_1x^{-1}$ and $\tD_1x^{-1}$
introduce a new order 
$x^{\delta -2}$ and also a contribution to the order $x^{\delta -1}$.

\subsubsection{General: $\tA_s x^{-s}$ and $\tD_s x^{-s}$}

More generally, the term $\tA_s x^{-s}$ introduces a lowest order $x^{\delta-s-1}$ linear in $\tA_s$. The following terms $\tA_{s+1}, \tA_{s+2}...$
also contribute to the order $x^{\delta-s-1}$, but there are negligible with respect to $\tA_s$. Furthermore, the previous terms $\tA_{s-1},
\tA_{s-2}...$ also appear at order $x^{\delta-s-1}$, but non-linearly. We can show from the observational constraints that we
have on the coefficients, that these non-linear contributions can be at most of the same order of magnitude as $\tA_s$, but not larger,
since they contain a term of the order $\epsilon^s$. Of course the same kind of terms are introduced by $\tD_s x^{-s}$.

Therefore, the equations at order $x^{\delta-s-1}$ ($s\geq 0$) are 
\bea
\label{eq:order_s_bis}
\lefteqn{\hat{F}^{(s)}_1(a_1,d_1)\cdot \tA_s+\hat{G}^{(s)}_1(a_1,d_1)\cdot \tD_s} \nonumber\\
&&=\hat{H}^{(s)}_1(\tA_1,..., \tA_{s-1}, \tD_1,..., \tD_{s-1},a_1,d_1)\nonumber\\
&&+\hat{Q}^{(s)}_1(\tA_{s+1},...,\tD_{s+1},...)\nonumber\\
\lefteqn{\hat{F}^{(s)}_2(a_1,d_1)\cdot \tA_s+\tilde{G}^{(s)}_2(a_1,d_1)\cdot \tD_s} \nonumber\\
&&= \hat{H}^{(s)}_2(\tA_1,..., \tA_{s-1}, \tD_1,..., \tD_{s-1},a_1,d_1)\nonumber\\
&&+\hat{Q}^{(s)}_2(\tA_{s+1},...,\tD_{s+1},...)\nonumber\\
\eea
where
\bea
\lefteqn{\hat{F}^{(s)}_1(a_1,d_1)=\sum_{i\in O_1}f_i(-1)^{\bi+\nui-1}2^{\gi+\ri-1}}\nonumber\\
&\times&\Big(2s\bi-s(s-1)\gi\Big)a_1^{\bi+\gi-1}d_1^{\mi+\nui+\ri}\,,\nonumber\\
\lefteqn{\hat{G}^{(s)}_1(a_1,d_1)=\sum_{i\in O_1}f_i(-1)^{\bi+\nui-1}2^{\gi+\ri-1}}\nonumber\\
&&\times\Big(-2\mi+2s\nui-s(s-1)\ri\Big)a_1^{\bi+\gi}d_1^{\mi+\nui+\ri-1}
\, .\nonumber\\
\eea

$\hat{H}_j^{(s)}$ can be at most of the order of magnitude of $\tA_s$ and $\tD_s$ whereas $\hat{Q}_j^{(s)}$ are much smaller since they
are of the order of magnitude of $\tA_{s+1}$ and $\tD_{s+1}$. 

A unique solution $(\tA_s, \tD_s)$ exists at order $s$ if 
\be
\label{eq:det2}
\det \left(\begin{array}{cc}\hat{F}^{(s)}_1(a_1,d_1)&\hat{G}^{(s)}_1(a_1,d_1)\\ \hat{F}^{(s)}_2(a_1,d_1)&\hat{G}^{(s)}_2(a_1,d_1) ) 
\end{array}\right)\neq 0~.
\ee
In this case, we can determine the solution at order $s$ as a function of the previous coefficients and the following coefficients.
Generically the contribution $\hat{H}^{(s)}_i$ from the previous coefficients is much larger than the one from the following 
coefficients $\hat{Q}^{(s)}_i$, since the constraints on the previous
coefficients is less stringent (see table \ref{table2}). We can therefore 
neglect the following coefficients and at each order determine the
solution as function of the previous coefficients. Nevertheless for some specific
case the contribution from the previous coefficients can become very small and even vanish. In this case, we have to take into account the
contribution of the following coefficients. This means that the solution $(\tA_s, \tD_s)$ will be of the order of magnitude of $(\tA_{s+1},
\tD_{s+1})$. At order $s+1$ we can then neglect $\tA_s$ and $\tD_s$ in
$\hat{H}^{(s+1)}_i$ which appear non-linearly and 
are therefore much smaller than the order $s+1$ linearly. This means that we can determine $(\tA_{s+1},\tD_{s+1})$ as a function
of the previous coefficients, but neglecting the order $s$. If the determinant (\ref{eq:det2}) is non zero for all $s\geq 0$,
we can find a solution order by order. On the other hand if the determinant vanishes for some $s$, the equations can lead to 
inconsistencies.

One remark has to be made about the order of magnitude of the coefficients. In the general case with four fields, we have found 
constraints on the coefficients $\tA_n, \tB_n, \tD_n$ and $\tE_n$ from the observations. Since the different fields are not related to the same
observation, the constraints are different for each field. We have therefore found that $\tD_n$ and $\tE_n$ can be much larger than $\tA_n$ and
$\tB_n$ for small $n$, see table (\ref{table2}). For large $n$ the situation is reversed. Nevertheless, from the previous analysis of the
equations, we see that this situation is not satisfying. Indeed, if two coefficients are much larger than the two others at order
$s$, it means that we can neglect the two small coefficients. Therefore, we introduce four new equations at order $s$ but only two new
coefficients and the determinant (\ref{eq:det2}) vanishes. Hence we
can not insure that a solution exists. Therefore, even if the
observations allow less stringent constraints on the order of
magnitude of some coefficients, the equations of motion generically
imply that all the coefficients can have at most the order of magnitude of the smallest one at each order.  
 
\subsection{Application to our problem}
\label{app:modification}

We need to transform the four equations of motion to apply the method
described above. First, we  multiply
each equation by the correct power of $e^{\nu}$ and $e^{\xi}$ such that each power in the equation (\ref{eq:simple_form})
is positive or null.

After this modification, equation (\ref{eq:constraint}) and (\ref{eq:t_r}) have the correct form and the method can be applied directly. For $n=1/2$, which is the case we consider in detail, equation (\ref{eq:r_rr}) reduces to 
\be\label{n1/2}
\left(\frac{\xi'+\nu'}{r}+\frac{\xi'^2}{4}+\frac{\xi'\nu'}{2}\right)\sqrt{\ck} =0 ~.
\ee

We can expand the two terms $f=\left(\frac{\xi'+\nu'}{r}+\frac{\xi'^2}{4}+\frac{\xi'\nu'}{2}\right)$ and $g=\sqrt{\ck}$ in power of $x$,
using the method described above. Let's call $\rho$ the lowest order of the development of $f$, containing only the coefficients
$a_1, b_1, d_1$ and $e_1$ and $\sigma$ the lowest order of $g$.

At lowest order the equation becomes 
\be\label{eq:ordre}
 f^{(\rho)} g^{(\sigma)}=0~.
 \ee

We have three possibilities:
\begin{itemize}
\item $f^{(\rho)}=0$ and $ g^{(\sigma)}\neq 0$.
The following order is then: $f^{(\rho+1)} g^{(\sigma)}=0$. Since $ g^{(\sigma)}\neq 0$, it implies $f^{(\rho+1)}=0$.
The same development can be made at each order, and therefore we find $f\equiv 0$.
\item $f^{(\rho)}\neq 0$ and $ g^{(\sigma)}= 0$.
As previously this implies $g\equiv0$. This case is the trivial case $\ck=0$ and is therefore not interesting. 
\item $f^{(\rho)}=g^{(\sigma)}= 0$.

In this case, the following order is $f^{(\rho+1)}g^{(\sigma+1)}=0$, which has exactly the same form as equation (\ref{eq:ordre}) and implies
therefore either $f\equiv 0$ or $g\equiv 0$.  

\end{itemize}

Since we want $\ck \neq 0$, equation (\ref{eq:r_rr}) becomes
\be
\label{eq:schwarz_1}
f=\left(\frac{\xi'+\nu'}{r}+\frac{\xi'^2}{4}+\frac{\xi'\nu'}{2}\right)=0
\ee
 
at each order. We recover one of the Schwarzschild equation for which the method can be applied easily.
Note that for the three equations (\ref{eq:constraint}), (\ref{eq:t_r}) and (\ref{eq:schwarz_1}),
the relations between the usual coefficients of Post-Newtonian parametrization are not modified by the additional coefficients
up to an order $10^{-9}$ which is the order of magnitude of the largest additional coefficient.

Equation (\ref{eq:t_tt}) is slightly different. Indeed, the left-hand side is proportional to $\ck^{n+1}$,
that means proportional to $\left(\frac{1}{Mr_s} \right)^{2(n+1)}$, whereas the right-hand side is proportional to
$\left(\frac{1}{Mr_s}\right)^2$. So the left-hand side is $\left(\frac{1}{Mr_s}\right)^{2n}\simeq (10^{23})^{2n}$ times larger
than the  right-hand side. Therefore, this equation can modify the relation between the usual coefficients. Indeed, we will now
mix the usual coefficients coming from the right-hand side, with the additional coefficients of the left-hand side which are
multiplied by $\left(\frac{1}{Mr_s}\right)^{2n}$.

We consider in the following the case $n=1/2$. The lowest order of the left-hand side is 8 whereas the lowest order of the right-hand
side is 5. At order 5 equation (\ref{eq:t_tt}) will then have the following form 
\be
\frac{1}{Mr_s\al}F(\tA_1,\tB_1,\tD_1,\tE_1,a_1,b_1,d_1,e_1)=G(a_1,b_1,c_1,d_1)~,
\ee

where $F$ is proportional to $\tA_1, \tB_1, \tD_1$ and $\tE_1$. This equation gives a relation between the usual and the additional coefficients.

Orders 6 and 7 have the same form, except that they contain also the
coefficients $a_2, b_2...$ and $\tD_0$. 
Order 8 is different. Indeed, at this order the left-hand side contains also $a_1, b_1, d_1$ and $e_1$.
Since they are multiplied by $10^{23}$, we can neglect the terms coming from the right-hand side.
The same occurs for all the following orders. Therefore, for orders 8
and larger, equation (\ref{eq:t_tt}) becomes 

\be
\left(\xi''+2\frac{\xi'}{r}+\frac{\xi'^2}{4}\right)\ck^{n+1} =0~.
\ee

The same argument as in Eq.~(\ref{n1/2}) implies that 

\be
\label{eq:schwarz_2}
\left(\xi''+2\frac{\xi'}{r}+\frac{\xi'^2}{4}\right)=0~.
\ee
This is the second Schwarzschild equation, but it is valid only for
orders larger than 7 in the development in power of $x$. 
For smaller orders we have to take into account the right-hand side.

The same situation occurs for powers smaller than 5. The left-hand
side can be neglected with respect to the right-hand side, 
and therefore we can apply the method to equation (\ref{eq:schwarz_2}).

To summarize, the method can be directly applied to equations (\ref{eq:constraint}), (\ref{eq:t_r}) and (\ref{eq:schwarz_1})
which replace (\ref{eq:r_rr}). Then we have to solve orders $x^5, x^6$
and $x^7$ of equation (\ref{eq:t_tt}). 
Our method can then be applied to equation (\ref{eq:schwarz_2}) for
orders larger than $x^7$ and smaller than $x^5$. 

\subsection{Solutions}
\label{app:solution}
Let us divide each equation by the correct power of $x$ such that the lowest order , containing only $a_1, b_1, d_1$ and $e_1$ is $x$ for each equation. Concerning equation (\ref{eq:t_tt}), this means that we have to divide by $x^6$.

At order $x$ for the equations (\ref{eq:constraint}), (\ref{eq:t_r}) and (\ref{eq:schwarz_1}) we find: 

\bea
b_1&=&-a_1\,,\nonumber\\
e_1&=&-\frac{a_1}{2}\,,\nonumber\\
0&=&(c_1+c_3)d_1\, .   \label{eq:d1}
\eea

Order $x$ of equation  (\ref{eq:t_tt}) will be treated separately since it contains also $\tA_1, \tB_1, \tD_1$ and $\tE_1$.

We impose $a_1=-1$ to recover Newton's theory and therefore we find  

\bea
\label{eq:ordre_1}
b_1&=&1\,, \nonumber\\
e_1&=&1/2\, .
\eea

From the last equality of equation (\ref{eq:d1}) we have  two possibilities: either $d_1=0$ or $c_3=-c_1$.

\subsubsection{$d_1=0$}

At order $x^2$ equation (\ref{eq:t_r}) implies 

\be-8(c_1+c_3)d_2=0 \,. \ee

Hence, either $c_3=-c_1$ or $d_2=0$. At order $x^3$ we have

\be
\Big[c_2-(c_1+c_3)\Big]d_3=0 .
\ee

Again we have two possibilities: either $c_2=c_1+c_3$, or $d_3=0$. The situation is the same at each order. Indeed if we assume that we
have chosen $d_1=d_2=...=d_{s-1}=0$ from the order $x$ to $x^{s-1}$, order $x^s$ gives 

\be
\label{eq:12_ds}
\Big[(c_1+c_2+c_3)s^2-3(c_1+c_2+c_3)s-2(c_1-c_2+c_3)\Big]d_s=0~.
\ee 

This means that the only possibility to have at least one of the $d_s\neq 0$, is to satisfy one of the relation 

\be
\label{eq:ds}
(c_1+c_2+c_3)s^2-3(c_1+c_2+c_3)s-2(c_1-c_2+c_3)=0~,
\ee

for some positive integer $s$. Therefore, we have to consider two situations: either $d_s=0+ O(\tD_0,\tE_0)$ for all $s$, or
one of the relation
(\ref{eq:ds}) is satisfied. 

In the first case, the usual parameters $d_s$ are equal to zero up to the order of magnitude of the additional coefficients
which of course modify
equation (\ref{eq:12_ds}). We can calculate the positive order $x, x^2...$ including the additional coefficients. And we have also the
negative order $x^0, x^{-1},...$. We can show that these sets of equations imply either 

\be
d_s\sim \tD_0~,~ \forall s \hspace{0.3cm}\textrm{and}\hspace{0.3cm} \tD_0\sim \tD_1 \sim \tD_2 ...\sim \tD_\infty
\ee

or 

\bea
d_1\sim \tE_0 d_2\sim \tE_0^2 d_3\sim ...\sim \tE_0^\infty d_\infty\nonumber\\
d_s\sim \tE_0 d_{s+1}\sim \tE_0^2 d_{s+2}\sim ...\sim \tE_0^\infty d_\infty\nonumber\\
\tD_0\sim \tE_0 d_1\sim \tE_0^\infty d_\infty\nonumber\\
\tD_s\sim \tE_0 \tD_{s-1}\sim ...\sim \tE_0^\infty d_\infty
\eea

Since $\tD_\infty\rightarrow 0$ in order that the expansion (\ref{eq:expansion_3}) converges in the Solar System, the first case implies
$A^r(r)=0$.

In the second case, since $\tE_0^\infty \rightarrow 0$ and $d_\infty$ is finite, we also have $A^r(r)=0$.

Hence $A^t(r)$ becomes the only degree of freedom of the vector
field, which is then completely determined by the constraint
(\ref{eq:constraint}).

In the following we will study in details the second situation where one of the $d_s$ at least is different from zero. We consider the simplest case where $d_1\neq 0$ and therefore $c_3=-c_1$.

\subsubsection{$c_3=-c_1$}
\label{app:sol_c1c3}

From order $x^2$ of the four equations (\ref{eq:constraint}),
 (\ref{eq:t_r}), (\ref{eq:schwarz_1}) and (\ref{eq:schwarz_2}),
and using eq. (\ref{eq:ordre_1}) we find

\bea
a_2&=&\frac{1}{2} \,, \nonumber\\
b_2&=&\frac{3}{8}\,, \nonumber\\
e_2&=&\frac{1}{8}+\frac{d_1^2}{2}\,, \nonumber\\
0&=&2c_2d_1^2-4c_2d_1+c_1-\frac{1}{2}c_2 \,.
\eea

We see that $b_1=1$ and $a_2=1/2$ are in complete agreement with the observations. The additional coefficients imply only
a contribution of order $10^{-9}$, which we have safely neglected since they are well beyond the precision
of the measurements. The last equation allows to calculate $d_1$ as a function of $c_1$ and $c_2$.

We have to determine the equations for orders $x^3$ and $x^4$ which
come from the mixed terms of equation (\ref{eq:t_tt}). We use the
solutions for the previous coefficients. 

From order $x^3$, we then find 

\bea
a_3&=&-\frac{3}{16}\,,\nonumber\\
b_3&=&\frac{1}{16}\,,\nonumber\\
e_3&=&\frac{1}{32}+\frac{3}{4}d_1^2+d_1d_2\,,\nonumber\\
d_3&=&\frac{1}{8c_2}\Big(-8d_2c_2-4c_1d_1-24c_1d_1^3+52d_1^3c_2\nonumber\\
&&+4c_1d_2+c_2d_1+32d_2c_2d_1^2\Big) ~.
\eea

From the order $x^4$ we obtain

\bea
a_4 &=&\frac{1}{16}\,,\nonumber\\
b_4 &=&\frac{1}{256}\,,\nonumber\\
e_4 &=& \frac{1}{128c_2}\Big(c_2+80c_2d_1^2+64d_1d_2c_2+64d_1c_1d_2\nonumber\\
&&+816c_2d_1^4-64c_1d_1^2-384c_1d_1^4+512d_2c_2d_1^3\nonumber\\
&& +64d_2^2c_2\Big)\nonumber\\
d_4&=&\frac{1}{192c_2^2}\Big(-104c_1d_2c_2+70c_1d_1c_2-208c_1d_1^3c_2
\nonumber\\
&&+16c_1^2d_2-16c_1^2d_1-32c_1d_2c_2d_1^2-96c_1^2d_1^3
\nonumber\\
&&-3264c_1d_1^5c_2 +132d_2c_2^2-652d_1^3c_2^2-7d_1c_2^2\nonumber\\
&&-384d_2c_2^2d_1^2+3840c_2^2d_1^4d_2+7104c_2^2d_1^5\nonumber\\
&&+384c_2^2d_1d_2^2\Big)\,.
\eea

We also need to solve for the negative powers. Using the hierarchy between the additional coefficients
$|\tA_1|\gg |\tA_2|...$ we find from equations (\ref{eq:constraint}), (\ref{eq:t_r}) and (\ref{eq:schwarz_1})
at order $x^0, x^{-1}$ and $x^{-2}$ that  

\bea
\tD_0 &=& -\frac{3\tB_2}{16d_1c_2}\Big(24c_2d_1^4-8c_1d_1^2+14c_2d_1^2-11c_2-2c_1\Big)\nonumber\\
\tA_1&=& \tB_2\,, \hspace{0.3cm}\tB_1=\frac{5\tB_3}{4}\,,
\hspace{0.3cm}\tE_1= -\frac{\tB_2}{2}\,, \hspace{0.3cm}\tD_1 =
\frac{3\tB_2}{4d_1}\,, \nonumber\\
\tA_2&=&-\tB_2\,, \hspace{0.3cm}\tE_2= \frac{\tB_2}{2}\,,
\hspace{0.3cm} \tD_2 =\frac{\tB_2c_1}{8c_2d_1}\, .
\eea

We remark that $\tB_1$ is of the order of $\tB_3$, therefore it can be neglected with respect to $\tA_1, \tD_1$ and $\tE_1$, but also with respect to the order 2. $\tB_2$ remains undetermined. $\tB_3$ will be determined by the lower order $x^{-3}$, but we are not interested in its value here.

We now consider equation (\ref{eq:t_tt}) at the order $x, x^0$ and
$x^{-1}$, which contains both the coefficients of negative and
positive powers. We solve this system of three equations plus the
equation (\ref{eq:d1}) with the help of maple. 
It contains the five variables $d_1, d_2, c_1, c_2$ and $\tB_2$ and
the two parameters $\alpha_1$ and $M$.
We find a set of solutions, from which we only consider those with
$c_1$ and $c_2$ negative in order to ensure a positive-definite
Hamiltonian for perturbations and non superluminal propagation of
spin-0 degrees of freedom  
in the approximately-Minkowski regime of the theory \cite{lim}

\bea
\label{eq:solc1_1}
c_1 &=&\frac{-33.11}{\alpha_1^2} \Big(\frac{ \tB_2}{Mr_s}\Big)^2\,,\hspace{0.7cm}
c_2 = \frac{-30.75}{\alpha_1^2}\Big(\frac{ \tB_2}{Mr_s}\Big)^2\,,\nonumber\\
d_1 &=&0.16\,,\hspace{2.3cm}d_2 = -0.62\,,
\eea

\bea
c_1 &=&\frac{-0.48}{\alpha_1^2} \Big(\frac{ \tB_2}{Mr_s}\Big)^2\,,\hspace{0.7cm} c_2 = \frac{-6.36}{\alpha_1^2}\Big(\frac{ \tB_2}{Mr_s}\Big)^2\,,\nonumber\\
d_1 &=&-0 .10\,, \hspace{2.3cm}d_2 = 0.28\,,
\eea

\bea
c_1 &=&\frac{-24.11}{\alpha_1^2} \Big(\frac{ \tB_2}{Mr_s}\Big)^2\,,
\hspace{0.7cm} c_2 = \frac{-143.48}{\alpha_1^2}\Big(\frac{
  \tB_2}{Mr_s}\Big)^2\,, \nonumber\\
d_1 &=&-0 .80\,, \hspace{2.3cm}d_2 = 0.72\,,
\eea

\bea
\label{eq:solc1_4}
c_1 &=&\frac{-140.29}{\alpha_1^2} \Big(\frac{ \tB_2}{Mr_s}\Big)^2\,,\hspace{0.7cm} c_2 = \frac{-61.93}{\alpha_1^2}\Big(\frac{ \tB_2}{Mr_s}\Big)^2\,,\nonumber\\
d_1 &=&0 .66\,,\hspace{2.3cm}d_2 = -7.89\,.
\eea

Each solution implies a strong constraint on the parameters $c_1, c_2$ and $\alpha_1$.
Indeed $\tB_2=B_2\left(\frac{r_s}{\rgm}\right)^2$ can be at most $10^{-32}$ in order not the be detectable
in the Solar System, and therefore $\Big(\frac{ \tB_2}{Mr_s}\Big)^2\simeq 10^{-18}$. This means that either $c_1$ and $c_2$
or $\alpha_1$ has to be very small if $\alpha_1 \neq 0$.

Finally, we have to calculate the determinant of the system to determine if
a solution exists at each order. Using the method described above we find from the positive powers in $x$
(which correspond to negative powers in $r$)($s>1$) 

\be
S\left (\begin{array}{c} a_s\\b_s\\d_s\\e_s \end{array}\right)
=\left (\begin{array}{c} H_1^{(s)} \\H_2^{(s)}\\H_3^{(s)}\\H_4^{(s)} \end{array}\right) \hspace{0.7cm}\textrm{where}
\ee

\be
S=\left (\begin{array}{cccc} -1&0&0&-2 \\0&s(s-1)&0&0\\-s&-s&0&0\\
\frac{d_1s(s-1)}{2}&0& -\frac{c_1(s-1)(s-2)}{4}&\frac{d_1s(s-1)}{2} \end{array}\right)
\ee

and $H^{(s)}_i$ are functions of the previous coefficients $a_1,...a_{s-1},b_1, ...,b_{s-1},d_1,..,d_{s-1},e_1,..,e_{s-1}$.
Therefore, at order $s$ a unique solution $(a_s,b_s,d_s,e_s)$ exists if 

\be
\det S=-\frac{c_1}{2}s^2(s-1)^2(s-2)\neq 0~.
\ee

Since $c_1\neq 0$, the condition is satisfied for $s>2$. The order $s=2$ for which the determinant vanishes has been
solved explicitly above and leads to no inconsistency. The solution is not unique, since we have found four different values
for $d_2$.
For each order ($s>2$) the determinant is non zero and therefore we can determine the unique solution as a function of the
previous orders.

The same structure repeats for positive powers of $r$ ($s>2$)

\be
\hat{S}\left (\begin{array}{c} \tA_s\\ \tB_s\\ \tD_s\\ \tE_s \end{array}\right)
=\left (\begin{array}{c} \hat{H}_1^{(s)}+ \hat{Q}_1^{(s)} \\
  \hat{H}_2^{(s)} +\hat{Q}_2^{(s)}\\ \hat{H}_3^{(s)}+\hat{Q}_3^{(s)}
\\ \hat{H}_4^{(s)}+\hat{Q}_4^{(s)} \end{array}\right) \hspace{0.7cm} 
\textrm{where}
\ee

\be
\hat{S}=\left (\begin{array}{cccc} 1&0&0&2 \\0&-s(s+1)&0&0\\s&s&0&0\\
\frac{c_1s(s+1)d_1}{2}&0& \begin{array}{c} c_2(14-s)d_1^2-\\
  \frac{c_1(s-1)(s-2)}{4}\end{array} &\frac{c_1s(s+1)d_1}{2}
\end{array}\right) 
\ee

and $\hat{H}^{(s)}_i$ are functions of $a_1$, $b_1$, $d_1$ and $e_1$, 
and of the previous coefficients $\tA_1,...,\tA_{s-1}$, $\tB_1,...,\tB_{s-1}$,
$\tD_1,...,\tD_{s-1}$,$\tE_1, ...,\tE_{s-1}$, and
$\hat{Q}^{(s)}_i\ll\hat{H}^{(s)}_i$. A unique solution exists at order $s$ if 
\be
\det \hat{S}=-2s^2(s+1)\Big(c_2(14-s)d_1^2-\frac{c_1}{4}(s-1)(s-2)\Big)\neq 0~.
\ee
The orders $s=0, 1$ and 2 have been calculated above.
For each of the solutions (\ref{eq:solc1_1}) to (\ref{eq:solc1_4}) we find that $\det \hat{S}\neq 0~\forall~ s$.

\twocolumngrid

\end{document}